\input{aipcheck}

\documentclass[
    ,final            % use final for the camera ready runs
  ]
  {aipproc}

\layoutstyle{6x9}

\begin{document}

\title{Physical origin of X-ray flares following GRBs}

\classification{95.30.Gv, 95.30.Lz, 95.85.Nv, 95.85.Pw, 98.70.Rz}
\keywords      {gamma-ray bursts; gamma-ray; X-ray; 
radiation mechanism; polarization}

\author{Bing Zhang}{
  address={Department of Physics, University of Nevada Las Vegas, Las
           Vegas, NV 89154}
}

\begin{abstract}
One of the major achievements of {\em Swift} is the discovery of the
erratic X-ray flares harboring nearly half of gamma-ray bursts (GRBs),
both for long-duration and short-duration categories, and both for
traditional hard GRBs and soft X-ray flashes (XRFs). Here I review the
arguments in support of the suggestion that they are powered by
reactivation of the GRB central engine, and that the emission site is
typically ``internal'', i.e. at a distance within the forward shock
front. The curvature effect that characterizes the decaying lightcurve
slope during the fading phase of the flares provides an important
clue. I will then discuss several suggestions to re-start the GRB
central engine and comment on how future observations may help to
unveil the physical origin of X-ray flares.
\end{abstract}

\maketitle

%%%%%%%%%%%%%%%%%%%%%%%%%%%%%%%%%%%%%%%%%%%%
%% MAINMATTER
%%%%%%%%%%%%%%%%%%%%%%%%%%%%%%%%%%%%%%%%%%%%

\section{Introduction}

The successful launch and operation of the {\em Swift} satellite
offers the opportunity to unveil the final gap between the prompt
emission and the late afterglow. With the co-operation of the Burst
Alert Telescope (BAT) and the X-Ray Telescope (XRT), a canonical early
X-ray afterglow lightcurve is
emerging\cite{nousek06,zhang06,chincarini05,obrien06}, which includes
five components beside the prompt emission itself (Fig.1), a steep decay
component, a shallower-than-normal decay component, a normal decay
component, a possible post-jet-break steep decay component, as well as
one or more erratic X-ray flares. Not every segment exists in every
burst, but all the lightcurves could be in principle understood within
such a general framework. These lightcurves bring invaluable
information to understand prompt emission - afterglow transition,
GRB emission site, central engine activity, forward-reverse shock
physics, and GRB immediate environment (for a full discussion, see
Ref. \cite{zhang06}).

\begin{figure}
  \includegraphics[height=.38\textheight, angle=-90]{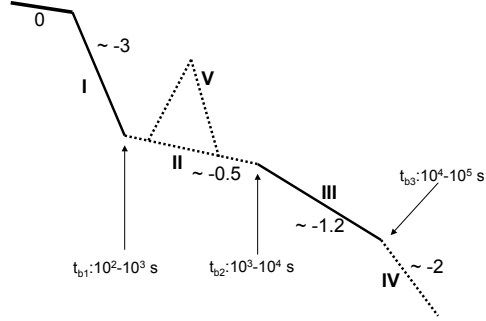}
  \caption{The canonical X-ray afterglow lightcurve. From Zhang et
   al. \cite{zhang06}.} 
\end{figure}

The most interesting component is erratic X-ray flares detected in
nearly half {\em Swift}
GRBs\cite{burrows05,falcone06,romano06,obrien06,liang06}. 
The general observational properties of these flares include
(e.g. \cite{burrows05b}): (1) flares typically have rapid rise and
fall times, with $\delta t/ t_{peak} \ll 1$; (2) many light curves
have evidence for an underlying afterglow power law component with the
same slope; (3) in many bursts, multiple early flares exist in a same
burst; (4) in some cases, e.g. GRB 050502B \cite{burrows05,falcone06},
the flux increases are very large (factors of tens to hundreds); (5)
flares soften as they progress; (6) the durations of the flares are
positively correlated with the epochs when the flares happen; (7) in
some cases, very late flares occur at around days after the trigger
(e.g. GRBs 050502B \cite{falcone06} and 050724
\cite{barthelmy05,campana06}; (8) there is no apparent difference
between long-duration and short-duration GRBs and between normal, hard
GRBs and soft XRFs as far as the X-ray flare properties are concerned.

\section{X-ray flare mechanism: late central engine activity and
internal energy dissipation} 

The data greatly constrained the possible models to interpret X-ray
flares. 
\begin{itemize}
\item The rapid rise and fall with $\delta t/ t_{peak} \ll 1$
strongly disfavor the scenarios that invoke large angular scale
external shock models\cite{zhang06,ioka05}, including refreshed shocks,
patchy-shell jets, multi-component jets, signatures induced by
neutron decay, etc.
\item The existence of multiple flares in a same GRB disfavors
scenarios that can only account for one flare, e.g. the synchrotron
self-Compton emission in the reverse shock\cite{kobayashi05}, the
deceleration of the blastwave\cite{piro05}, and the progenitor models
invoking a companion\cite{macfadyen05}.
\item The comparable fluence (and hence, energy budget) of the giant
X-ray flare of GRB 050502B\cite{burrows05,falcone06} disfavors models
that interpret X-ray flares and prompt emission without additional
energy input, including the clumpy medium model\cite{zhang06}.
\item The similar flaring behavior observed in both long and short
bursts suggest that the flare mechanism is insensitive to the
progenitor\cite{perna06}. 
\end{itemize}

The leading model is that flares are due to the reactivation of the
GRB central engine, and that they are produced from an ``internal''
dissipation radius within the external blastwave
front\cite{burrows05,zhang06,ioka05,fw05}. There are two advantages
for such a model to interpret the flare data\cite{zhang06}. First,
re-starting the GRB central engine effectively re-sets the
starting-time. For example, should there be no prompt gamma-rays, the
onset of a late flare would be effectively the ``trigger'' time. Each
episode of the late central engine activity is equivalent to each
other and should define its own time zero point ($t_0$). Such a shift
is essential to define the most relevant temporal decay index in the
$\log F_\nu - \log (t-t_0)$ lightcurves, especially when $t$ is not
much larger than $t_0$ (here all the numerical number of the time
quantities are with respect to the trigger time). If $t_0$ is far from
the trigger time, the conventional lightcurves ($\log F_\nu - \log t$,
with time zero point at the trigger time) would show artificial very
steep decays as observed (sometimes the decay index after the flare
peak is steeper than -7). Second, this model is very economical
energetically. This is because we invoke two different emission sites
to interpret the X-ray flares and the background power-law decaying
X-ray afterglow. The former are from an internal radius, while the
latter is from the external forward shock. In order to produce the
same X-ray flux level, the internal model usually only requires a
small amount of energy budget with respect to that of the prompt
emission, and the 
lightcurve including the prompt emission and the flares essentially
reflects the time history of the central engine energy input. For the
external shock model, on the other hand, one requires at least
comparable energy budget with the intial blastwave in order to make
any noticeable change of the afterglow flux level\cite{zm02}. The
flare amplitudes are usually a factor of several to several hundreds. 
This would require an unbelievably large energy input from the central
engine without any observational signature. This is rather implausible. 
For external shock models that invoke very small density
clumps\cite{dermer06}, the energy budget problem could be eased, but
so far no modeling could successfully reproduce the giant flares as
observed in GRB 050502B.

Invoking an unknown central engine activity inevitably introduces
difficulties to model the flares (as compared with the simply
afterglow model). However, one has a clean signature that could be
used to diagnose the above interpretation. This is the so-called
``curvature effect'' that dominates the decaying phase of the
flare. Assuming that radiation stops abruptly at an internal radius
from a conical shell, and that the cooling frequency is below the
X-ray band before the cessation of the synchrotron emission, no fresh
electrons would contribute to the emission in the X-ray band since the
cessation of the emitter. What one detects is then
the emission from higher latitudes with respect to the viewing
direction that propagates to the observer with slight delay due to the
extra distance the photons need to travel. There is a clear prediction
for such a scenario: the temporal decay index is connected to the
spectral index by $\alpha=\beta+2$ with the convention $F_\nu \propto
\nu^{-\beta} t^{-\alpha}$
\cite{kumar00,dermer04,zhang06,fw05,panaitescu06,dyks06}. Such a
prediction is rather robust regardless of the speed history of the
shell\cite{zhang06,fw05} and is largely insensitive to the jet
structure as long as the viewing angle is not very far away from the 
bright core of the structured jets\cite{dyks06}.

The clean $\alpha=\beta+2$ characteristic is complicated by two
additional effects \cite{zhang06}. One is the $t_0$ effect, i.e. one
needs to identify the correct central engine starting time for the
flare in investigation. If the internal-origin of the flares is
correct, this $t_0$ should be at the beginning of the rising segment
of the flare. The second effect is the superposition effect. There is
an underlying afterglow level which tends to make the decay slope
shallower, especially near the transition time between the steep decay
component (segment I in Fig.1) and the shallow decay components from
the afterglow (segments II or III in Fig.1). However, since both the
spectral index ($\beta$) and the temporal decay indices in different
segments ($\alpha_1$, $\alpha_2$ and $\alpha_3$) could be directly
inferred from the data, one has a clean strategy to test the
correctness of the internal-curvature interpretation.

Liang et al. \cite{liang06} have performed such a systematic test. 
The procedure is the following. (1) Identify a steep decay component
with $\alpha>2$ following a certain X-ray flare (or the steep decay
component following the prompt emission). (2) Perform a fit to the
afterglow lightcurve with two overlapping power laws. The late-time
index is fitted by the data and the early-time index is assumed to be
$\beta+2$. (3) Search for the time zero point $t_0$ that gives the
best fit to the data. The results are impressive. In many cases, the
searched $t_0$ is exactly where it is expected, i.e. at the beginning
of the corresponding flare (Fig.2). This gives strong support to our
beginning hypothesis: that the flares are from internal dissipation of
energy and that the decay is controlled by the curvature effect. In
some cases the signature is not as clear as expected, but these cases
also have overlapping flares or extended emission so that the rising
part of the flare is usually burried beneath other emission
components. 

One interesting thing to note is that both GRB 050502B (a
long-duration burst \cite{falcone06}) and GRB 050724 (a short-duration
burst associated with non-star-forming galaxy \cite{barthelmy05}) 
have a late flare peaking around a day. The decay after the peak
clearly follows the prediction of curvature effect, strongly
suggesting that the central engine is turned on again (for an extended
emission period) at such a late epoch. This poses great challenge to
the possible central engine models for both long and short GRBs.

%%%%%%%%%%%%%%%%%%%%%%%%%%%%%%%%%%%%%%%%%%%%
%% Sample figure:
%%
%% The option [height=...] scales the picture to the given height,
%% without it it would be printed at its nominal size
%%%%%%%%%%%%%%%%%%%%%%%%%%%%%%%%%%%%%%%%%%%%

\begin{figure}
\centering
\hspace{-2cm}
\begin{minipage}[t]{.35\textwidth}
  \includegraphics[height=.20\textheight]{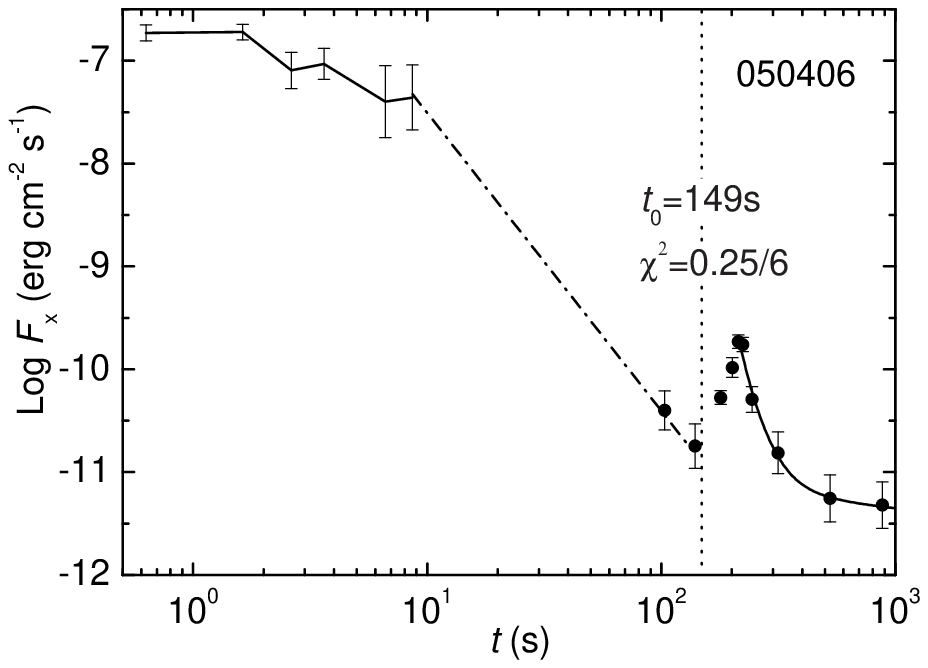}
  \includegraphics[height=.20\textheight]{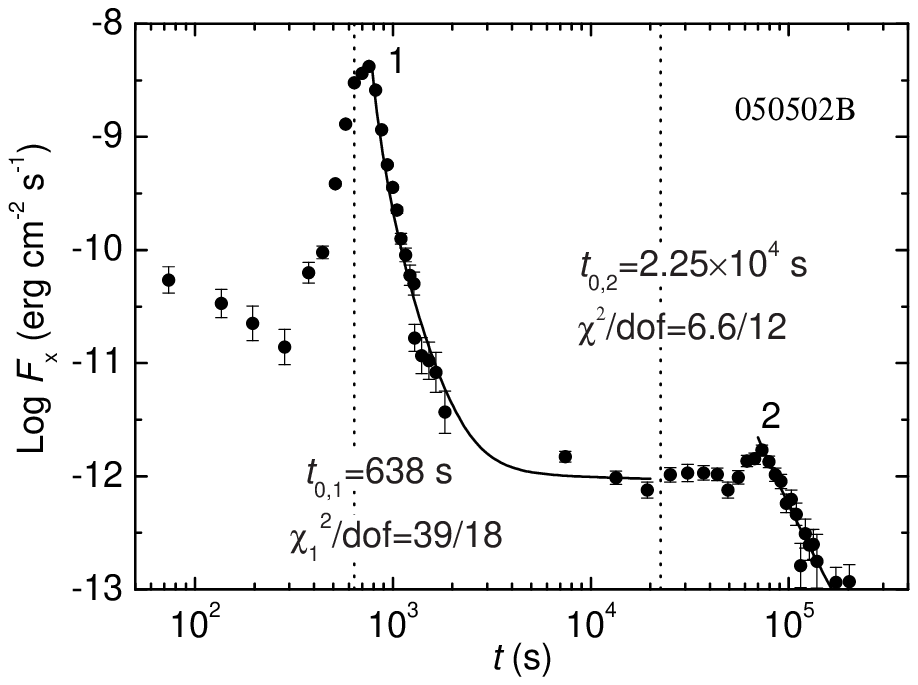}
\end{minipage}
\hspace{1.5cm}
\begin{minipage}[t]{.35\textwidth}
  \includegraphics[height=.20\textheight]{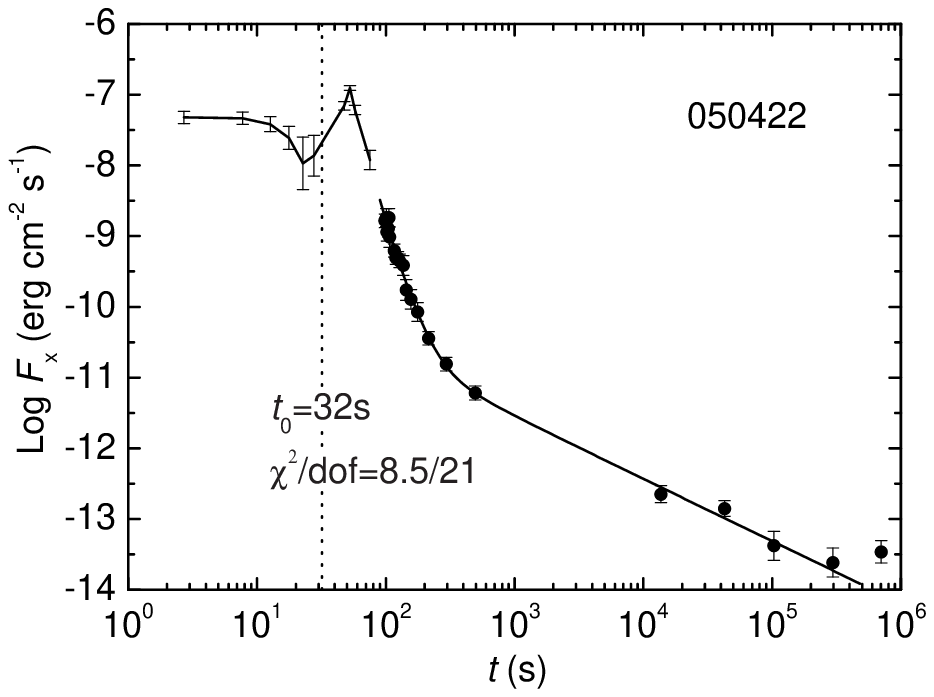} 
  \includegraphics[height=.20\textheight]{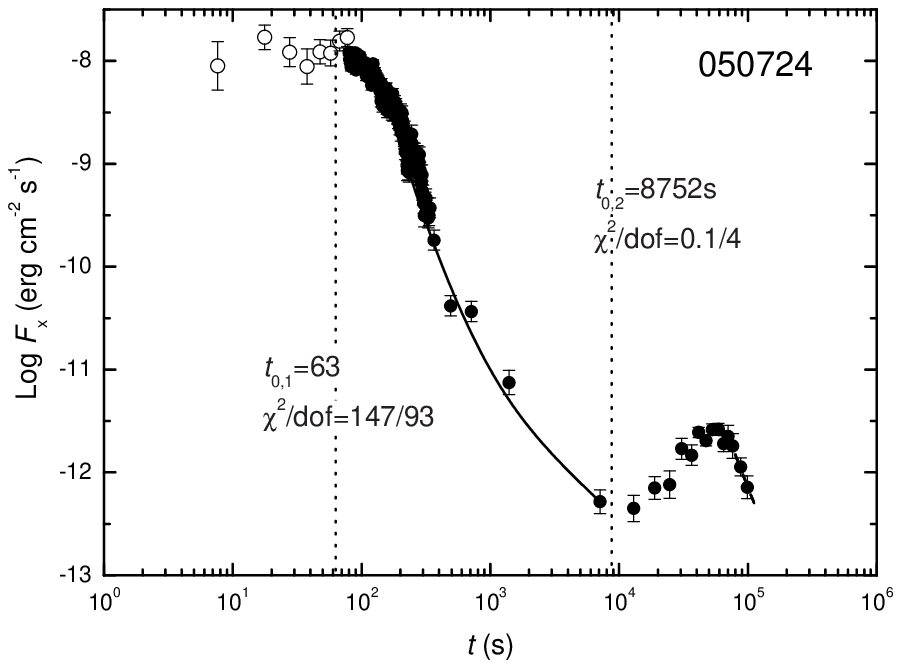}
\end{minipage}
  \caption{Several examples of the searched $t_0$ of X-ray flares.
   From Liang et al. \cite{liang06}.}
\end{figure}

\section{How to restart the central engine}

Now that the X-ray flare data suggest that the GRB central engine must
be re-started at a delayed time and that it works in a non-continuous
manner, a straightforward question is then how a GRB central engine
model would be able to give rise to what is observed.

This is still largely an open issue. Nonetheless a list of suggestions
have been made since the discovery of the X-ray flares.

\begin{itemize}
\item King et al. \cite{king05} suggest that within the collapsar
progenitor model for long GRBs, the rotating stellar core may undergo
fragmentation during the collapse. The accretion of the fragmented
blobs into the newly formed black hole offers a plausible mechanism of
X-ray flares. Whether such a fragmentation can be realized calls for
proofs from numerical simulations. The existence of similar X-ray
flares in short GRB 050724 \cite{barthelmy05} suggests that the flare
mechanism likely also work for progenitors without a heavy envelope.
This makes this suggestion inconclusive.
\item Based on energetics alone, Fan et al. \cite{fan05} argue that
the flares following the short GRB 050724 (and possibly also GRB 050709)
must be of magnetic origin. This is because in various compact star
merger models, the mass of the fuel to power the accretion is limited
to be at most about 1 solar mass. The durations of the early flares
typically last for hundreds seconds, suggesting that the accretion
rate ($\leq 0.01 {\rm M_\odot/s}$) is too low for the
neutrino-annihilation mechanism to power the observed $10^{48} ~{\rm
erg ~s^{-1}}$ luminosity. As a result, the X-ray flares are expected
to be somewhat linearly polarized, although the detailed degree of
polarization is hard to predict. The argument may be also extended to
flares in long GRBs
\item Motivated by the similarity of the X-ray flare properties for
both long and short bursts, Perna et al. \cite{perna06} argue that the
flaring mechanism must be related to the property of something in
common for both types of bursts. They suggest the accretion disk is
this common link, and the episodical flares may be caused by
fragmentation of the accretion disk itself. Gravitational instability
in the outer parts of the disk may cause the fragmentation. The model 
can naturally interpret some properties of the flares, including the
duration-timescale correlation and the duration-peak luminosity
anticorrelation. 
\item Based on results of previous MHD numerical simulations and
theoretical analyses, Proga \& Zhang \cite{proga06} suggest that the
accretion flow does not have to be chopped at large radii due to
fragmentation. Rather, strong magnetic fields near the black hole can
play the role to modulate the accretion flow. The magnetic barrier,
which has been evident in previous numerical simulations
\cite{proga03,ina03}, can act as an agent to repeatedly stop and
restart the accretion flow. This gives a plausible mechanism to power
X-ray flares in both long and short GRBs. In this model, the launched
jet is magnetized, which coincides the argument based on energetics
\cite{fan05}. 
\item Dai et al. \cite{dai06} argue that the postmerger product of
double neutron star merger may well be a massive neutron star with
milliseocnd rotation period. The differentially-rotating stellar body
would wind up interior poloidal magnetic fields to form progressively
stronger toroidal magnetic fields. These tangled fields then float up
and break through the stellar surfaces, triggering
magnetic-reconnection-driven explosive events. This is an attractive
mechanism to produce X-ray flares in short GRBs. A less specified
central engine model with the similar idea is discussed in \cite{gao06}. 
\end{itemize}

Entering the second-year operation, {\em Swift} will keep collecting
flare data for a large sample of bursts. From data point of view, it
is high time to systematically analyze the flare data and perform
statistical analyses of the properties of flares. 
In the mean time, detailed theoretical modeling of the flare
properties (e.g. \cite{wu06}) would shed more light on the radiation
mechanism and the dissipation site of X-ray flares. In long terms,
coordinated observations of {\em Swift} and high energy detectors such
as {\em GLAST} would reveal whether there are high-energy counterparts
of X-ray flares, and hence, greatly constraint the emission physics of
the flares \cite{wang06}. An even far-reaching goal is to detect
polarization of X-ray flares with future X-ray polarimeters, which
would lead to more direct diagnoses of the GRB central engine
\cite{fan05}. 

\begin{theacknowledgments}
I thank stimulative collaborations with E. W. Liang, Y. Z. Fan,
J. Dyks, D. N. Burrows, S. Kobayashi, P. M\'esz\'aros, R. Perna,
P. J. Armitage, D. Proga, Z. G. Dai, X. Y. Wang, X. F. Wu, J. Nousek,
N. Gehrels and many other members of the {\em Swift} team on various topics
covered in this talk. This work is supported by NASA through grants
NNG05GB67G, NNG05GH91G, and NNG05GH92G.
\end{theacknowledgments}

%%%%%%%%%%%%%%%%%%%%%%%%%%%%%%%%%%%%%%%%%%%
%% The following lines show an example how to produce a bibliography
%% without the help of the BibTeX program. This could be used instead
%% of the above.
%%%%%%%%%%%%%%%%%%%%%%%%%%%%%%%%%%%%%%%%%%%

\end{document}